# Interaction between graphene and $SiO_2$ surface

X.F. Fan[a,*], W.T. Zheng[a], Z.X. Shen[b] and Jer-Lai Kuo[c]

a. College of Materials Science and Engineering, Jilin University, Changchun 130012, China

b. Division of Physics and Applied Physics, Nanyang Technological University, Singapore 637371

c. Institute of Atomic and Molecular Sciences, Academia Sinica, Taipei 106, Taiwan

(2011-7-15)

**Abstract**

With first-principles DFT calculations, the interaction between graphene and $SiO_2$ surface has been analyzed by constructing the different configurations based on α-quartz and cristobalite structures. The single layer graphene can stay stably on $SiO_2$ surface is explained based on the general consideration of configuration structures of $SiO_2$ surface. It is also found that the oxygen defect in $SiO_2$ surface can shift the Fermi level of graphene down which opens out the mechanism of hole-doping effect of graphene absorbed on $SiO_2$ surface observed in experiments.

**Introduction**

Graphene, which is a monolayer of carbon atoms with honeycomb lattice[1], has attracted enormous attention[2] due to the fascinating physical properties[3], such as abnormal quantum Hall effects[4] and massless Dirac fermions[5], which are ascribed to the liner dispersion near Dirac points in the *k* space. In additional, with extremely high mobility of carriers, graphene is expected to be the kernel material in the next generation of carbon-based nanoelectronics[6-8]. However, the prominent electronic properties of graphene are very sensitive to the change of external conditions[8-10]. The electronic states near Dirac points are modulated easily due to the absorption of some molecules[11, 12], structural corrugation[13, 14] and interaction with substrate surfaces[15, 16]. Obviously, this issue is critical to the wider electronic-device applications, such as field effect transistors. With the view of two-dimensional materials as a transport layer in the switching devices, the interaction between graphene with the dielectric layer is important. The change of structural and electronic properties of graphene due to the interaction deserves to be analyzed.

In present, two surfaces (SiC and $SiO_2$) have drawn a great attention. Graphene can be grown epitaxially on SiC substrate by high temperature annealing. Therefore, the interaction between graphene and SiC surface have been largely investigated[16-19]. The other one is $SiO_2$ surface which is the thin oxide layer of Si substrate and widely used as an insulating medium for Si-based device design[4, 5, 8, 9, 20]. After the graphene is discovered on $SiO_2$ surface by optical measurement, various experiments about graphene sheets and nanoribbons[21], including the device fabrication and the measurement of fundamental properties, have been taken on $SiO_2$ substrate[4, 5, 8, 9, 20]. The properties of graphene absorbed on $SiO_2$ are possible to be modulated has been reported by many experiments[4, 8, 22, 23]. The stable gate-controlled conduction has proven that the few-layer graphene can be stable on $SiO_2$ surface[20]. In the seed paper of Novoselov et al., the electronic measurement showed that single layer graphene on $SiO_2$ surface is possible to be doped chemically[5]. In the experiment of Tomero et al., the graphene was found to be initially p-type and changed to be n-type after annealing at 200℃ in vacuum for the graphene devices fabricated on $Si/SiO_2$ substrate[24]. With scanning probe microscopy, Ishigami et al. considered that graphene sheet just partially conformed to the $SiO_2$ substrate[22]. The Raman spectroscopy observed the spatially inhomogeneous doping effect of graphene supported by $SiO_2$ substrate[23]. By anneal experimental, thermal annealing can induce increased coupling between graphene and $SiO_2$ surface and the absorbed oxygen is activated to accept the charge from graphene[25, 26]. The theoretical calculation will play an important role to elucidate the micromechanism of interface between graphene and $SiO_2$ and help to explain the different phenomena in the experiments.

---
[*] E-mail: xffan@jlu.edu.cn

Kang et al. considered that graphene stayed on O-polar, Si-polar and the partially hydrogenated Si-polar α-quartz (0001) surface[27]. They found that free-standing graphene could stay on Si-terminated surface. The electronic properties of graphene near Dirac points were modified obviously on O-polar and the partially hydrogenated Si-polar surface. Shemella et al. analyzed O-polar surface and found that the π electronic properties of graphene were destroyed fully[28]. With hydrogen termination, graphene can freely stand on the surface. Hossain et al. found that graphene was adsorbed on specific sites of O-terminated surface and the charge transferred from graphene to the O (or Si)-terminated surface[29]. Interestingly, Nguyen et al. found the O-termination surface could be reconstructed and thus graphene could freely stand on O-polar surface[30]. The linear dispersion of π electrons was retained with a small band gap at the Dirac point and the charge transfer didn't found. Evidently, though many theoretical calculations have been performed to consider the interaction between graphene and $SiO_2$ surfaces, the results from different theoretical studies seem to be not consistent fully. There is not an explicit conclusion that can give a clear physical insight into the interaction and explain the experimental phenomena.

In this work, we analyze the interaction between graphene and $SiO_2$ surface by constructing the models based on α-quartz and cristobalite structures. $SiO_2$ surface from $Si/SiO_2$ involves a numerous number of configurations due to the amorphous nature. This results in that the atomic detail at the interface between graphene and $SiO_2$ is difficult to simulate. Thus, in order to consider the interaction at the interface, some typical local atomic configurations need to analyze to elucidate the micro-mechanism of interaction. In the previous theoretical study, the different configurations were not considered in detail. We systematically considered the different cases for the interaction, including the Si-polar surface with two dangling bonds per surface Si atom, Si-polar with one dangling bond per surface Si atom, O-polar surface, Si-adsorbed surface, O-adsorbed surface, reconstructed O-polar surface and surface with defects. Based on the analysis of different configurations, we explain the reason why single layer graphene can stay stably on $SiO_2$ surface and the charge distribution of the graphene absorbed on $SiO_2$ is inhomogeneous. We also find that the oxygen defect in $SiO_2$ surface results in the hole-doping phenomenon of graphene absorbed on $SiO_2$ surface.

**Structures and Method**
**1. Model Structures**
The interaction between graphene and $SiO_2$ surfaces is simulated using the repeated-slab model. In order to avoid the spurious vertical coupling effect, the vacuum separation is set to be more than 15 Å. In the present work, both structures (α-quartz and cristobalite) are used to construct the surface of $SiO_2$. For α-quartz structure, (0001) surface is chosen to simulate the interaction. Seven layers of silicon dioxide with H-passivated bottom surface are used in order to eliminate the effect of interaction of both surfaces. For the 1×1 surface of α-quartz, the 2×2 cell of graphene can be matched properly with small difference of lattice constants. Thus, both oxygen (O)-terminated and silicon (Si)-terminated 1×1 surfaces are used to model the interaction with graphene. In order to simulate the case of oxygen (or silicon) absorbed $SiO_2$ surface, the 2×2 supercell is constructed to interact with one layer of graphene with 4×4 cell (32 carbon atoms). In order to simulate other configurations of $SiO_2$ surface, the (111) surface of cristobalite is adopted to model the interaction with graphene layer, though cristobalite is a high-temperature polymorph of $SiO_2$. As same as the α-quartz (0001) surface, the lattice constant of 1×1 surface can match properly with that of 2×2 cell of graphene. Twelve layers of silicon dioxide are used to eliminate the coupling effect of both surfaces. 2×2 surface is also used to simulate O- (or Si-, OH-, and SiH-) absorbed $SiO_2$ surface.

**2. Calculation Details**
In this work, all the calculations have been performed with density functional theory[31]. The generalized gradient approximation (GGA) is used to express the exchange-correlation energy of interacting electrons by the parametrization of Perdew-Burke-Ernzerhof (PBE)[32]. With the accurate frozen-core full-potential projector augmented wave (PAW) method[33], the electron-core interaction is described as implemented in the VASP program package[34, 35]. The *k*-space integral and plane-wave basis, as detailed below, are tested to ensure that the total energy is converged at the 1 meV/atom level. The kinetic energy cutoff of 600 eV for the plane wave expansion is found to be sufficient. For 1×1 surface, the geometry is optimized with an 8×8×2 *k* mesh. 2×2 surface is constructed based on the optimized 1×1 surface and then is optimized with a 2×2×1 *k* mesh. The self-consistent electronic structure calculations are performed with a 12×12×1 *k* mesh for 1×1 surface and a 6×6×1 *k* mesh for 2×2 surface.

**Results and Discussions**

## 1. Electronic Properties of Bulk and Surface

Before launching the calculation of surface, the electronic properties of bulk are investigated to uncover the structural and electronic properties. The calculated lattice constants *a* (4.848 Å) and *c* (5.371 Å) are similar to the experimental lattice parameters (4.913 Å and 5.405 Å), for α-quartz structure with symmetry group P3121. For cristobalite structure with symmetry group FD-3M, the obtained lattice constant is 7.363 Å. By the consideration of [111] direction, the hexagonal cell with lattice constants *a* (5.20 Å) and *c* (12.75 Å) are employed. After the optimization of structure, the electronic properties of α quartz and cristobalite structures are obtained. The calculated bands are plotted versus high-symmetry lines in hexagon-symmetry Brillouin zone. As shown in Fig. 1, cristobalite structure and α-quartz have the similar band gap. Large band gaps demonstrate their insulating property which is consistent with the previous calculation[36, 37]. The difference is that the valence band edge and conduction band edge of cristobalite is at the same point (Γ point). The valence band edge of α-quartz is not at the Γ point. As mentioned in the previous study[38], the band gap is underestimated by DFT-GGA calculation. The exact estimation needs to perform the quasiparticle band structure calculation[38]. However, this doesn't affect the analysis about the electronic properties of the mixed system of graphene and $SiO_2$ surface, as stated in the next part.

It is well known that the $SiO_2$ surface in most experiments is with amorphous structure. For the amorphous $SiO_2$, the bond lengths and bond angles have continuous distribution[36], without the periodic cell which makes the difficulty to simulate the surface with DFT. However, the local structures of some crystal forms of $SiO_2$ are similar to that of amorphous $SiO_2$. The crystal $SiO_2$ structures, such as α quartz and cristobalite, have local structures of fourfold tetrahedral bonding for Si and twofold bridging bonding for O. For α quartz, the calculated Si-O bond length and the bond angle of Si-O-Si are about 1.590 Å and 143.05°, respectively. For cristobalite, the bond length and the bond angle are about 1.594 Å and 180°, respectively. The similar local structure results in the similar electronic insulating property. In addition, the different bond angle of α quartz and cristobalite can make the different surface configurations. Therefore, it is proper to use the surfaces of α quartz and cristobalite to simulate that of amorphous $SiO_2$, since the local structures of amorphous surface with different dangling bonds can be simulated by the different surfaces of α quartz and cristobalite.

Now we analyze some basic configurations of $SiO_2$ surface. If the top surface is the oxygen or hydroxyl (OH), the Si atoms of bottom surface are passivated by hydrogen. If the top surface is silicon or SiH, the oxygen atoms of bottom surface are also passivated by hydrogen. In Fig.2, seven configurations of surface are demonstrated, based on α-quartz and cristobalite. For Si-polar surface as a local stable configuration, the Si atom is possible to have two dangle bonds (Fig. 2A) and one dangle bond (Fig. 2B and C), or no dangle bond if silicon atom is passivated by two hydrogen atom. For O-polar surface as a local stable configuration, the oxygen atom is possible to have one dangle bond (Fig. 2F), or no dangle bond if oxygen atom is passivated by one hydrogen atom. It is noticed that each oxygen pair is bonded with one silicon atom (Fig. 2D) for α-quartz (0001) surface as a simple cleaved surface. The oxygen pair may be unstable and reconstruct the surface by removing one of oxygen atoms (Fig. 2G) or be passivated by partly hydrogenating (Fig. 2E) or fully hydrogenating.

Before investigating the interaction between graphene and different configurations, we calculate the electronic properties of different surfaces. As shown in Fig. 3A, we can found that the two dangling bonds of Si atom introduce two localized bands in the region of bang gap. Since there are two isolated electrons for silicon on the surface, one of the bands is occupied fully and leave one vacant band near the conduction band edge. Therefore, it is possible to be not activated. If one electron is paired by introduction of hydrogen, the residual electron will form a typical dangle bond. The case is similar to the Si atom in the cristobalite (111) surface. There is a localized state in the center of band gap and the state is half occupied (Fig. 3B and C). In Fig. 3D, we show the band structure of O-polar surface with one oxygen pair. It is well known that each oxygen atom has an isolated electron. Two isolated electrons from two oxygen atoms interact with each other and form two separated bands in the center of band gap and one of bands is occupied. However, this case may be different from that of Si atom of surface in Fig. 3A, since the two electrons don't form a stable pair due to the large distance. If one of oxygen atoms is passivated by hydrogen, we can find that the localized state is shifted to the top of valance band and is half occupied (Fig. 3E). This case has a little different from the case of cristobalite (111) surface. As shown in Fig. 3F, there are two localized bands which are almost degenerate.

## 2. Absorption of Graphene on Different Surface Configurations

We first consider the graphene layer is absorbed on the O-polar surface. Fig. 4A-E shows the optimized geometric structures for different adsorption sites of surface. It is found that graphene layer keeps its plane with hexagonal network for all these cases with physical absorption, except the chemical absorption site in the potential curve (Fig. 4G) for pair oxygen configuration (Fig. 4B). For the reconstructed surface (Fig 4A), the result is similar to the previous experimental and theoretical studies [30, 39, 40]. The difference is that the absorption energy is smaller (about 12 meV/unit cell). From the equilibrium position of potential curve in Fig. 4F, we can found that the distance between carbon and oxygen is about 3.8 Å. It demonstrates that the reconstructed surface is very stable and the interaction between graphene and the surface is weak, as Nguyen et al. stated[30]. There is no charge transfer between graphene and the reconstructed surface and the surface just disturbs weakly the electronic state of graphene at Dirac points to result in a small gap which even can be ignored. For (111) cristobalite surface, the distance between carbon and oxygen (about 3.0 Å) is similar at different absorption sites (Fig. C, D and E). The hexagonal center is the stablest site with absorption energy about 470 meV per unit cell, as shown in Fig. 4H. For top site of (0001) α-quartz surface with pair oxygen, the physical absorption is with equilibrium distance 3.4 Å and absorption energy 112 meV per unit cell. However, after hopping a potential barrier about 260 meV, the chemical absorption is formed with the distance between carbon and oxygen 1.55 Å, as shown in Fig. 4G. For the chemical absorption, the carbon has a trend to drop down to form a chemical bond with oxygen atom of surface. The absorbed carbon atom deviates from the graphene plane to result in large strain energy due to the localization of absorption site. As a case, the chemical absorption at top site of (111) cristobalite surface is possible to form. However, the absorption is not stable due to the large strain energy (Fig. 4H). The same reason is for partly hydrogenated O-polar surface on which graphene just form the physical absorption. Of course, for the fully hydrogenated O-polar surface, the interaction between graphene and surface is weak, since there is no active site for chemical absorption or strong physical absorption as other author stated.

For the Si-polar surface, the optimized geometric structures for different adsorption sites of surface are shown in the Fig. 5A-E. For all the case, graphene is absorbed physically on the surface and graphene layer keeps its planar hexagonal network. The binding energy is smaller than that of the absorption on O-polar surface. For (0001) α-quartz surface, the absorption energy is about 13 meV per unit cell with equilibrium distance about 4.3 Å. For (111) cristobalite surface, the binding energies (top and bridge: 18.5 meV/unit cell; hexagonal center: 24 meV/unit cell) are little larger than that of (0001) α-quartz surface. This may be attributed to the activity of silicon atom with one dangling bond. For both surface, the hexagonal center is the stablest site, similar to the case of O-polar surface. The reason is that the charge repulsion between graphene and surface of $SiO_2$ is weak due to the space separation of carbons and surface atoms. In additional, on top site of (111) cristobalite surface, graphene is possible to form chemical absorption due to the activity of dangling bond of silicon atom, as shown in Fig. 5H. This case is similar to the configuration of silicon with one dangling bond on (0001) α-quartz surface due to H-termination, as stated by Kang et al.. Of course, for the fully hydrogenated Si-polar surface, the absorption is physical. In general, as the above discussion suggests, the graphene layer can be absorbed freely on $SiO_2$, except the special case. The interaction between both them is weak and the electronic properties of graphene are possible to be changed weakly.

## 3. Electronic properties of interface without/with defects

In Fig. 6, the band structures of some typical configuration of graphene absorbed on surface are demonstrated. As Fig. 6A shown, when graphene layer is absorbed physically on the O-polar (0001) α-quartz surface with pair oxygen, the band structure of graphene is not changed obviously. It is found that the vacant band from pair is occupied partly and the Fermi level is pinned at that band and electron is transferred from graphene layer to surface to result in the p-doping of graphene. If the graphene is absorbed chemically on the surface after jumping the energy barrier, the band structure of graphene is broken strongly with a large band gap (about 2.7 eV), as previously reported[27]. If there is isolated oxygen on the surface, the localized band from oxygen will result in the shift of Fermi level, as shown in Fig. 6B. The band structure of graphene is not changed when it absorbed on bridge site of O-polar (111) cristobalite surface. The results from the top and hexagonal sites are similar to that of bridge site. For Si-polar surface, there is no charge transfer between graphene and surface, whatever there is the nonactive Si atom with two free electrons or the active Si atom with single dangle bond on the surface. This may be attributed to the similar electronegativity of silicon on $SiO_2$ surface and carbon of graphene. If the surface (O-polar or Si-polar) is passivated by hydrogen, the bands from $\pi$ electrons of graphene are mostly stayed in the large band gap of the surface and the electronic properties of graphene are not affected from the surface, as shown in Fig. 6C and F.

From the above analysis, the interaction between graphene and surface mostly is by physical absorption. The weak interaction doesn't break the band structure of graphene as reported in a lot of experiments. However, it is also possible to form the chemical absorption on special configuration, such as the oxygen pair configuration on O-polar surface. But previous studies showed that cleaved $SiO_2$ is easy to be reconstructed or hydroxylated (or hydrogenated) to eliminate the dangle bonds. Therefore, it seems to be impossible that there is large region of active surface with dangle bonds. While the oxygen and silicon defects with dangle bond is possible to appear on surface.

In Fig. 7, we give an analysis of some typical $SiO_2$ surfaces with defects. For oxygen defects (Fig. 7A and B) and silicon defect with one dangle bond (Fig. 7D) in the H-passivated (O-polar or Si-polar) surface, graphene layer still be absorbed physically on the surface and the active defect atom doesn't interact chemically with the carbon atom of graphene. However, the oxygen defects introduce the localized energy levels at top of valance band and result in that the Fermi level of graphene is shifted down. The charge transfer evidently induces the hole-doping of graphene. For silicon defect, there is no charge transfer and the electronic properties don't change. The defect just induces a localized level at the Dirac points. However, the electronic scattering from the defect will affect the mobility of carrier in graphene layer. Therefore, the defects result in the spatially inhomogeneous distribution of charge in graphene plane. In some special case, it is possible to appear the raised defects on the surface. In Fig. 7C and E, such the oxygen and silicon defects are demonstrated, respectively. It is found that the raised defects can form the chemical interaction with the graphene layer by dropping carbon atom down to form the C-O bond or C-Si bond. Such the defects disturb strongly the electronic properties of graphene near Dirac points and a band gap is opened obviously. The oxygen defect results in down-shift of the Fermi level of graphene and the silicon defect results in up-shift of the Fermi level. One energy level from defect is coupled with the bands from Dirac cone and the Dirac cone is broken near the coupling region.

**Conclusion**

The interaction between graphene and $SiO_2$ surface has been studied systematically by first-principles DFT calculations. Different configurations of $SiO_2$ surface are considered, based on α-quartz and cristobalite structures. It is found that graphene layer mostly is absorbed physically on surface and the hexagonal center site is the stablest site for both O-polar and Si-polar surface. This may be attributed to the largest spacial separation of charges to reduce the coulomb repulsion when graphene layer approaches to the region of surface due to van der Waals interactions and possible charge transfer. It is found that the absorption on O-polar surface is stronger than that on Si-polar surface. It is ascribed to the effect of charge transfer on O-polar surface which increases the electrostatic interaction between graphene and the surface. Due to the charge transfer, it is found that the graphene absorbed on O-polar surface is hole-doped, as observed on the experiments.

The H-passivated surfaces don't affect the electronic properties of graphene whatever there is O-polar surface or Si-polar surface, whereas the surface with defects will have the strong effect to the electronic properties of graphene. It is found that the oxygen defect in the surface can result in the hole-doping of graphene. The raised atom defects will interact chemically with graphene layer and disturb the band structure of π electrons. The raised oxygen results in down-shift of Fermi level of graphene and the raised silicon results in up-shift of Fermi level. Therefore, we suggest that $SiO_2$ surface which is used as a substrate for fabrication of graphene device should be cleaved with atomic level of planeness and then is passivated by hydrogen.

Fig.1

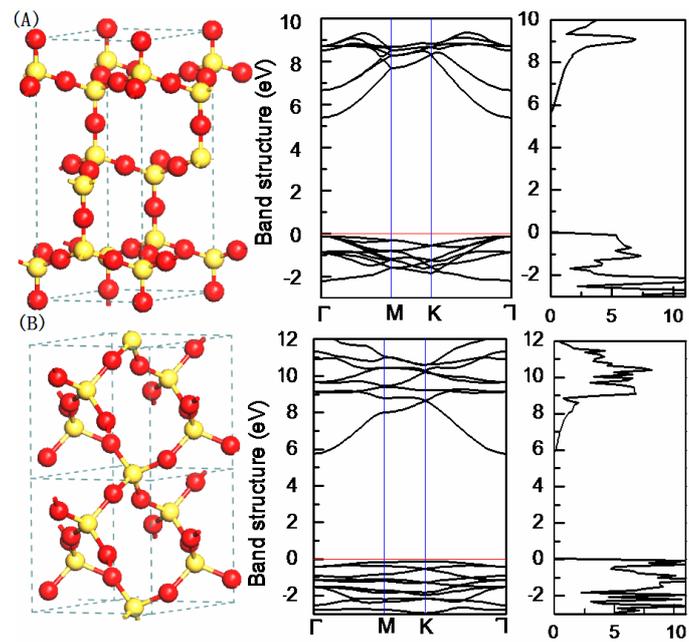

Fig. 1 The schematic structure, band structure and density of states of cristobalite (A) and α-quartz (B) structures

Fig. 2

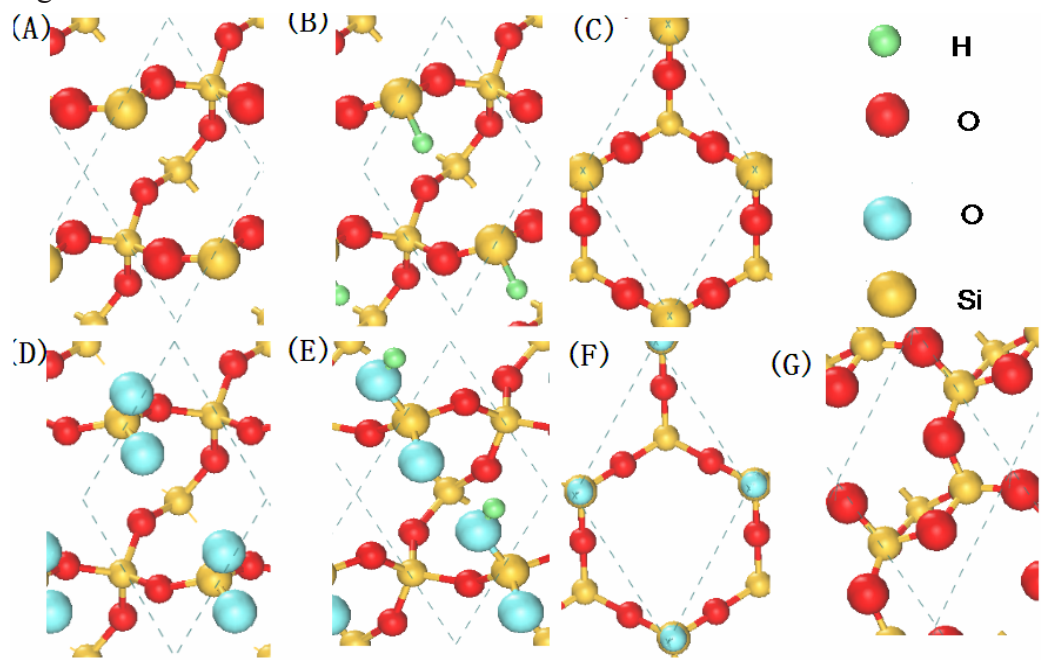

Fig.2 Schematic representation of the structures of seven basic $SiO_2$ surface: Si-polar (0001) α-quartz surface(A), Si-polar (0001) α-quartz surface with one dangling bonds per surface Si atom (B), Si-polar (111) cristobalite surface (C), O-polar (0001) α-quartz surface (D), O-polar (0001) α-quartz surface with one H-terminated oxygen (E), O-polar (111) cristobalite surface (F) and reconstructed O-polar (0001) α-quartz surface(G) .

Fig.3

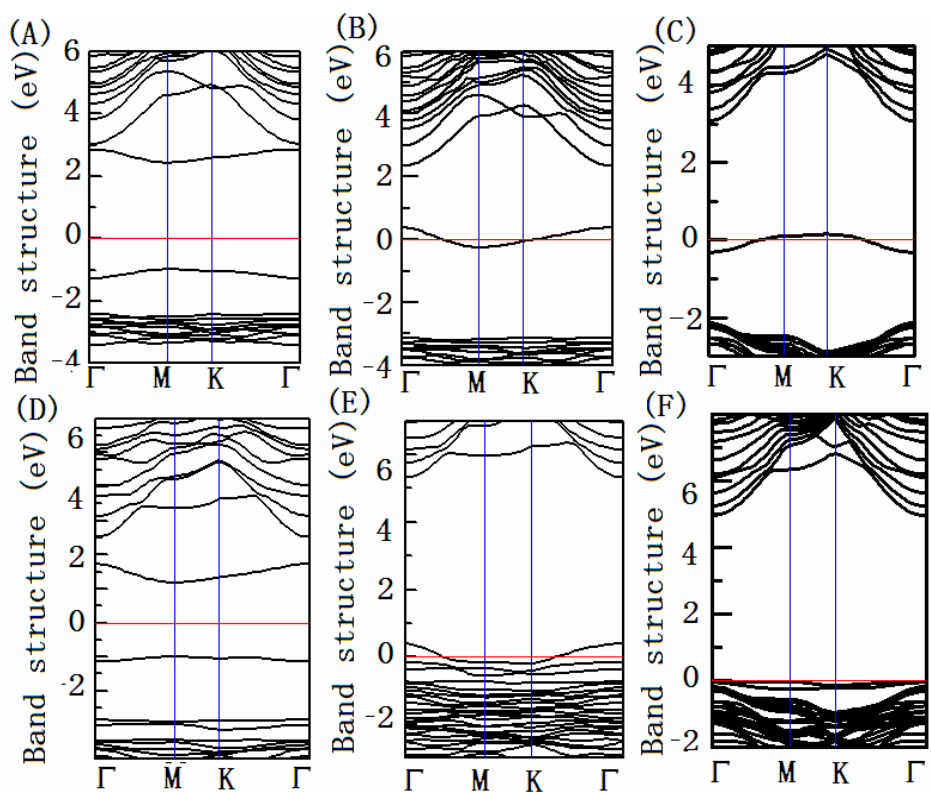

Fig.3 The band structures of the surface models of α-quartz and that of cristobalite: Si-polar (0001) α-quartz surface(A), Si-polar (0001) α-quartz surface with one dangling bonds per surface Si atom (B), Si-polar (111) cristobalite surface (C), O-polar (0001) α-quartz surface (D), O-polar (0001) α-quartz surface with one H-terminated oxygen (E) and O-polar (111) cristobalite surface (F).

Fig. 4

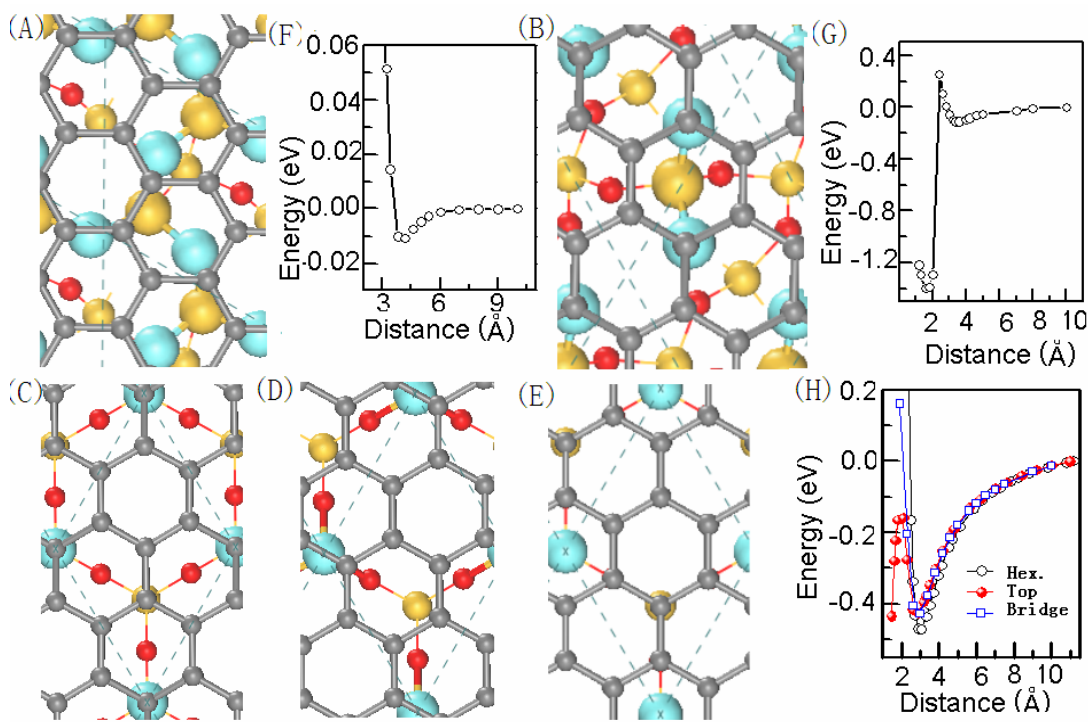

Fig. 4 Top views of optimized structures of graphene on O-polar surface including bridge site of reconstructed α-quartz surface (A), top site of (0001) α-quartz surface with pair oxygen and top (C), bridge (D) and hexagonal center sites (E) of (111) cristobalite surface, and potential-energy curves as a function of the oxygen atom of surface and the nearest-neighbour carbon atom of graphene including the curve (F) of configuration (A), curve (G) of configuration (B) and curves (H) of configurations (C), (D) and (E). The surface unit cell is shown by the dashed lines.

Fig. 5

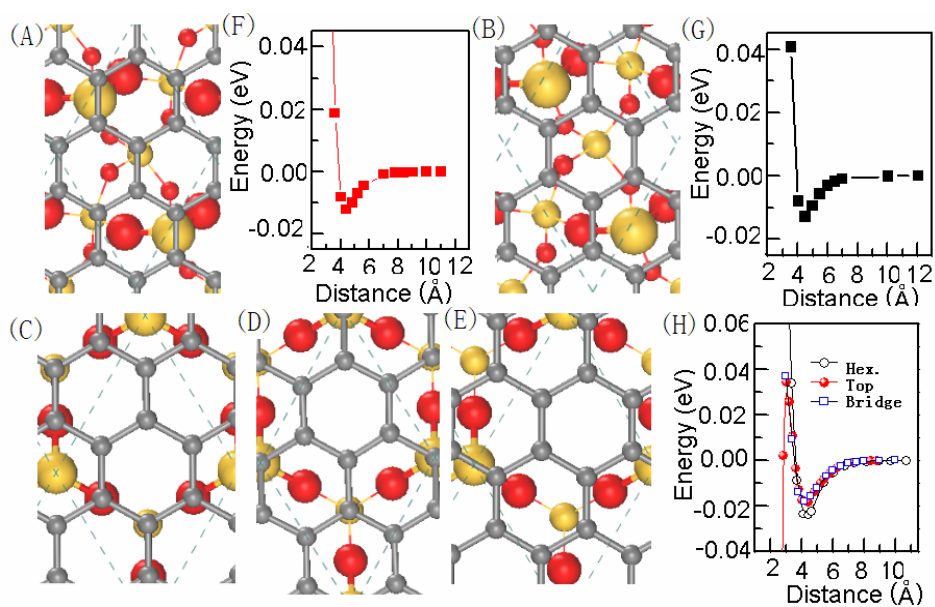

Fig. 5 Top views of optimized structures of graphene on Si-polar surface including bridge (A) and hexagonal center sites of (0001) α-quartz surface and hexagonal center (C), top (D) and bridge (E) sites of (111) cristobalite surface, and potential-energy curves as a function of the oxygen atom of surface and the nearest-neighbour carbon atom of graphene including the curve (F) of configuration (A), curve (G) of configuration (B) and curves (H) of configurations (C), (D) and (E). The surface unit cell is shown by the dashed lines.

Fig.6

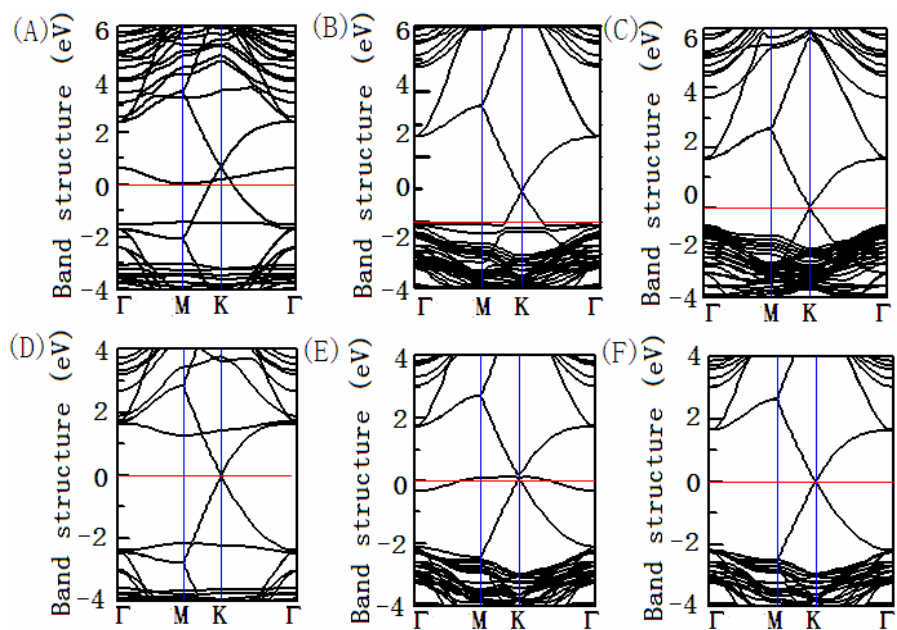

Fig. 6 The band structures of the mixed systems of graphene absorbed on $SiO_2$ surface: O-polar (0001) α-quartz surface with pair oxygen due to physical absorption on top site (A), O-polar (111) cristobalite surface with bridge site (B), H-terminated O-polar (111) cristobalite surface with bridge site (C), Si-polar (0001) α-quartz surface with bridge site (D), Si-polar (111) cristobalite surface with hexagonal center site (E) and H-terminated Si-polar (111) cristobalite surface with top site.

Fig. 7

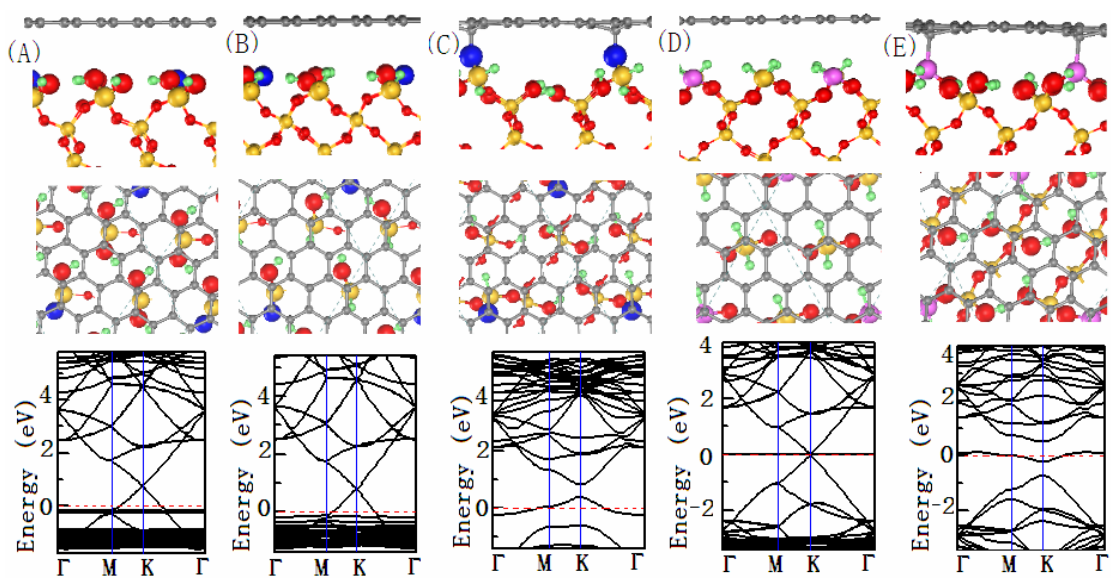

Fig. 7 Side and top views of optimized structures of graphene adsorbed on (0001) α-quartz surface with defect and the band structure of the mixed system including the OH surface with one oxygen (A), OH and SiH mixed surface with one oxygen (B), OH surface with one H-Si-O defect (C), $SiH_2$ surface with SiH defect (D) and OH surface with SiH defect (E). Blue sphere is for oxygen atom as a defect and pink sphere is for silicon atom as a defect.